\documentclass[11pt]{article}
\usepackage{amsfonts}
\usepackage{pstricks}
\usepackage{pst-node}
\usepackage{amsmath}
\usepackage{graphics}
\usepackage{graphicx}

\usepackage{epsfig}
\usepackage{subfig}
\usepackage{amssymb}

\begin{document}

\def\ba{\begin{eqnarray}}
\def\ea{\end{eqnarray}}
\def\w{\wedge}

\begin{titlepage}
\title{\bf Anisotropic cosmological solutions  to the $Y(R)F^2$ gravity}
\author{\"{O}zcan Sert$^{a}$\footnote{\tt osert@pau.edu.tr} \hskip 0.5cm and \hskip 0.3cm Muzaffer Adak$^{b}$\footnote{\tt madak@pau.edu.tr}\\ \\
 {\small $^a$Department of Mathematics, Faculty of Arts and Sciences, Pamukkale University}\\
 {\small 20017 Denizli, Turkey }\\
 {\small $^b$Department of Physics, Faculty of Arts and Sciences, Pamukkale University}\\
 {\small 20017 Denizli, Turkey } }
  \vskip 1cm

\maketitle

 \thispagestyle{empty}

\begin{abstract}
 \noindent
We investigate  anisotropic cosmological solutions  of   the theory with 
non-minimal  couplings between electromagnetic fields and
gravity in $Y(R) F^2$ form. After we derive  the field equations  by the
variational principle, we look for spatially flat cosmological solutions with magnetic fields  or electric fields. Then we give exact anisotropic  solutions by assuming the hyperbolic expansion functions. We observe  that the solutions   approach to the  isotropic case  in late-times.
\vskip 1cm


\end{abstract}
\end{titlepage}

\section{Introduction}

\noindent


The recent observations on the accelerated expansion of the universe \cite{Spergel2003,Peiris2003,Spergel2007,Perlmutter1999,Knop2003,Amanullah2010,Riess1998,Astier2006,Riess2007,Weinberg2013,Schwarz2016} lead to re-questioning of Einstein's General Relativity  and  searching  of new theories of gravitation.
Firstly, it was considered that the acceleration  may be caused by a constant energy arising from empty space
  and then  Einstein's field equations were  modified by adding a cosmological constant \cite{Weinberg1989}.
  This modification also   solved another problem of General Relativity  which called  dark matter \cite{Overduin2004,Baer2015} by  the model $\Lambda $CDM.
  But the observed value of the cosmological constant in this  model is not consistent with the large value of the zero-point energy which  calculated by quantum field theory. In addition, there are some observational discordance to overcome \cite{Ade2016,Hildebrandt2016,Jong2015,Kuijken2015,Fenech2017,Riess2011,Riess2016,Ade2014} in $\Lambda $CDM model.
  Moreover it revealed  some conceptual problems,
 such as what is the cosmological constant fundamentally? 
  
  To overcome these challenges and answer this question  theoretically, the following two scenarios were mostly considered.  
  The first is that it may be caused by a particular phase of   a dark energy field (quintessence) which  is a varying unknown energy source  or a scalar field fills the universe \cite{Peebles2003,Padmanabhan2003,Tsujikawa2013}. Secondly  the cosmological constant effects can be obtained by modifying   gravity and then gravity behaves very differently than  Einstein's theory of gravitation on the extremely large scales.
Among the modified  theories,  mostly $f(R)$  theory was  investigated  in literature (for reviews see \cite{Nojiri2011,Sotiriou2010,Capozziello2010}).

On the other hand, gravitating systems such as galaxies, galaxy clusters,
stars and  planets have also intrinsic magnetic fields \cite{Kim1990,Kim1991,Clarke2001,
Giovannini2004,Widrow2002,Grasso2001}. In order to take into account  the effects of electromagnetic fields to gravity, we have to consider the  Einstein-Maxwell theory for obtaining an exact description. Since the Einstein's theory  is modified to explain the dark matter and dark energy, then the Einstein-Maxwell theory can also be modified in the non-minimal form, especially, in the presence of the extreme situations with very strong fields such as the beginning of the universe.


    Such a non-minimal modification with $RF^2$-type was firstly considered  by Prasanna  \cite{Prasanna1971} to understand the complex nature between curvature and electromagnetic fields.  Then  the charge conservation was generalized to the  such terms \cite{Horndeski1976}.   The non-minimal couplings also can appear from
 dimensional reduction of   Gauss-Bonnet gravity \cite{Muller-Hoissen1988} and $R^2$  gravity \cite{Buchdahl1979,Dereli1990}.
  It is remarkable that the non-minimal modifications with $RF^2$ form  can arise from the calculation of QED
    one-loop vacuum polarization on a curved background \cite{Drummond1980}.
   Furthermore, the non-minimal $RF^2$ couplings may generate the primordial magnetic fields by quantum fluctuations at the inflation \cite{Turner1988,Lambiase2004}, and  the more general $R^nF^2$ couplings  can increase the amplitude to sufficiently large seed fields which lead to the present galactic magnetic fields \cite{Mazzitelli1995,Campanelli2008,Lambiase2008}. 
    In other words, the couplings can lead to quantum fluctuations of electromagnetic fields  at the inflationary stage by breaking the conformal invariance  \cite{Turner1988,Lambiase2004,Mazzitelli1995,Campanelli2008,Lambiase2008,Bamba2008}.
    Because of the inflation, the  scale of the  fluctuations can be
    stretched towards outside the Hubble horizon. Thus, they can
    be the reason of the large scale magnetic fields  observed
    in  galaxies.

 Since the electromagnetic fields lead to  anisotropic energy-momentum tensor and pressure without the averaged field assumption,  the inflation may be explained by using  the anisotropic Bianchi-I space-times  in the non-minimal $Y(R)F^2$ theory \cite{AADS2017}.  The more general modifications   have the spherically symmetric static solutions to explain the dark matter effects \cite{Sert2011EPJC,Sert2011MPLA,Sert2012Plus,Sert2013MPLA}, the regular black hole solutions to avoid the central singularity \cite{Sert2016regular} and  the pp-wave solutions \cite{Sert2011PRD}. It is interesting to note that the stability of the anisotropic Bianchi-I  solutions were investigated   for the extended $I(\phi,R,X)F^2$ theory recently in \cite{Do2017}. Furthermore, dark energy models with the non-minimally massive vector field couplings to gravity 
  have also been investigated in \cite{Bohmer2007}.

Therefore the non-minimal cosmological models need to more   investigation, in particular,  to explain the  late-time acceleration together with the origin of cosmic magnetic
fields  and its role in the evolution of the universe. Therefore,
 in this paper, we look for new anisotropic cosmological solutions with the hyperbolic expansion functions
  to  the $Y(R)F^2$ gravity and determine the corresponding models.

\section{The theory of $Y(R)F^2$ gravity} \label{model}

\bigskip
\noindent We obtain the field equations by varying an action
$I=\int_M{L}$ where $L$ denotes a Lagrange 4-form, and $M$ a
four-dimensional differentiable and orientable manifold endowed
with a metric $g = \eta_{ab} e^a \otimes e^b$,
$\eta_{ab}=\mbox{diag}(-+++)$. We set the orientation by the Hodge
star $*1 = e^0 \wedge e^1 \wedge e^2 \wedge e^3$. Here $e^a$ figures the
orthonormal basis 1-form. The Cartan-Maurer structure equations
 \begin{eqnarray}
 & &T^a = de^a + {\omega^a}_b \wedge e^b \, , \\
 & &{R^a}_b = d{\omega^a}_b + {\omega^a}_c \wedge {\omega^c}_b \, 
 \end{eqnarray}
define the torsion 2-form, and the curvature 2-form, respectively,
where $\omega_{ab}=-\omega_{ba}$ is the metric compatible connection
1-form. Since we will force to vanish the torsion,  $\omega_{ab}$ will be the Levi-Civita connection 1-form.

We consider the following Lagrangian for the non-minimal $Y(R)F^2$ gravity  
 \begin{eqnarray}\label{lag1}
  L =  \frac{1}{2\kappa^2} R*1 -Y(R) F\wedge *F + \lambda_a  \wedge  T^a \;.
   \end{eqnarray}
  Here the form of the non-minimal function $Y(R)$ will be determined by solutions and $\kappa$ is a gravitational constant, $R$ is the Ricci  curvature
scalar, $F=dA$  is the electromagnetic field  2-form,   $\lambda_a$  is the Lagrange
multiplier 2-form constraining  torsion to zero
($T^a=0$). We adhere the following shorthand
notations throughout the paper: $ e^a \wedge e^b \wedge \cdots =
e^{ab\cdots}$, \ $\iota_aF =F_a, \ \iota_{ba} F =F_{ab}, $ \ $
\iota_a {R^a}_b =R_b,   \ \iota_{ba} R^{ab}= R $ where $\iota$
denotes the interior product such that $\iota_b e^a = \delta^a_b$ and $\iota_a \iota_b ... = \iota_{ab...}.$
The independent variations of the Lagrangian with respect to $e^a$,
${\omega^a}_b$ and $F$ yield (up to a closed form)
 \begin{eqnarray}\label{generaleinsteinfe1}
   \delta{L}  &=& \frac{1}{2 \kappa^2} \delta{e}^a \wedge R^{bc} \wedge *e_{abc}
     + \delta{e}^a \wedge Y(R) (\iota_a F \wedge *F - F \wedge \iota_a *F) + \delta{e}^a \wedge D \lambda_a \nonumber \\
  & & +  \delta{e}^a \wedge  2Y_R (\iota_a R^b)\iota_b(F \wedge *F)
      + \frac{1}{2} \delta\omega_{ab} \wedge ( e^b \wedge \lambda^a - e^a \wedge \lambda^b) \nonumber \\
 & & + \delta{\omega}_{ab} \wedge  {\Sigma}^{ab} -\delta{ F} \wedge 2 Y(R)  *F   + \delta{\lambda}_a \wedge T^a
\end{eqnarray}
where  $Y_R =\frac{dY}{dR} $ and $\Sigma^{ab}$ is the angular momentum
3-form
\begin{eqnarray}\label{sigmaab1}
 {\Sigma}^{ab} &=&  D   [\iota^{ab}( Y_R F  \wedge *F )].
   \end{eqnarray}
We can solve $\lambda_a$ from $\delta \omega_{ab}$-equation
\begin{eqnarray}\label{lambdaaeb2}
 \lambda^a &=&  2\iota_b   {\Sigma}^{ab}  +\frac{1}{2} (\iota_{bc} {\Sigma}^{bc})\wedge e^a.
\end{eqnarray}
After the substitution of $\lambda_a$ into $\delta e^a$-equation
and some simplifications we arrive at the modified Einstein's
equation
\begin{eqnarray}\label{gfe1}
\frac{1}{2 \kappa^2}  R^{bc} \wedge *e_{abc} +   Y(\iota_a F \wedge *F - F \wedge \iota_a *F)
+ 2 Y_R (\iota_a R^b)\iota_b( F \wedge *F ) & & \nonumber \\
+   D [ \iota^b d(Y_R F_{mn} F^{mn} )]\wedge *e_{ab} &=&0   ,
\end{eqnarray}
while the modified Maxwell equations read
 \begin{eqnarray}
    dF=0 \, , \label{maxwell1} \  \hskip 2 cm
    d(Y*F)=0 \, . 
 \end{eqnarray}
where  $ d(Y_R F_{mn} F^{mn} ) =D(Y_R F_{mn} F^{mn} )  $   and it can be shown that $ 2 Y_R (\iota_a R^b)\iota_b( F \wedge *F )  = Y_RF_{mn}F^{mn} *R^a$.  In order to avoid the difficulties and instabilities of the last term in   the gravitational field equation (\ref{gfe1}), 
we continue with the condition
\begin{eqnarray}\label{cond1}
Y_R F_{mn} F^{mn} = - \frac{1}{\kappa^2}
 \end{eqnarray}
  where the constant $-\frac{1}{\kappa^2}$ is determined by the trace of the gravitational field equation
 (see \cite{AADS2017} for a detailed discussion). It is worthwhile to note that the condition (9) is not a new equation. Actually it corresponds to the conservation of energy-momentum tensor. If we take the exterior covariant derivative of the field equation (\ref{gfe2}) we obtain the condition again \cite{Sert2017}. Furthermore, the condition (9) causes  the  Ricci scalar $R$ to be  dynamic and it relates the electromagnetic field with the derivative of the non-minimal function $Y(R)$. 
 Here  $\kappa $  is the  gravitational coupling constant and  it determines the strength of the coupling  between electromagnetic and gravitational fields. Then the gravitational field equation can be written as 
 \begin{eqnarray}\label{gfe2}
-\frac{1}{2 }  R^{bc} \wedge *e_{abc} = \kappa^2 \tau_a\;,
\end{eqnarray}
where the effective energy momentum tensor $\tau_a= T_{ab} *e^b$ is
\begin{eqnarray}
\tau_a =    Y(\iota_a F \wedge *F - F \wedge \iota_a *F)
-\frac{1}{\kappa^2}*R^a \;.
\end{eqnarray}
Thus the effective energy density and pressures are given by $\rho= T_{00} $, $p_x = T_{11}$, $p_y=  T_{22}$, $p_z = T_{33}$\;.

 \section{Cosmological  solutions}

\bigskip

\noindent 

We look for cosmological solutions in the presence of  electromagnetic fields to the modified fields equations (\ref{gfe1}) and (\ref{maxwell1}) to describe  the evolution of the universe.  Since there is a preferential direction along  the electromagnetic field,  the energy-momentum tensor and the space-time metric become anisotropic unless assuming the  averaged  electromagnetic fields \cite{Tolman1930}.
 Then we consider the anisotropic,
   locally rotationally  symmetric Bianchi-I  metric
\begin{equation}\label{metric}
              g = - dt^2  +  a(t)^{2}dx^2 + b(t)^2 ( dy^2 + dz^2)
\end{equation}
where $a(t)$ is the expansion function in the $x$ direction and  $b(t)$ is  the planar expansion function in the $y$ and $z$ directions.    In order to obtain compatible solutions  with the above geometry (\ref{metric}) with rotational symmetry around the x-axis,  we choose  the electromagnetic field along the x direction.
 \begin{eqnarray}\label{electromagnetic1}
   F =E(t)e^{01} + B(t) e^{23}
 \end{eqnarray}
  with the electric component  $E(t)$ and   magnetic component   $B(t)$. 
  Here $e^0=dt$, $e^1=a(t) dx$,   $e^2=b(t) dy$, and $e^3=b(t) dz$
are the orthonormal basis 1-forms. Then we can set $B(t)=0$ or $E(t)=0$ to obtain solutions for  the sub cases with only electric fields or magnetic fields.  While the non-zero magnetic field is effective  at later times,   the non-zero electric field  is more important at the beginning of the universe, in the charged plasma. 
 
 Then the modified Maxwell field equations (\ref{maxwell1}) give
 \begin{eqnarray}\label{EB}
    B=\frac{B_0}{b^2} \;,\hskip 2 cm E = \frac{E_0}{Y b^2}
 \end{eqnarray}
where $B_0$ and $E_0$ are integration constants. On the other hand,
the modified Einstein field equation (\ref{gfe2}) yields 
\begin{eqnarray}
\frac{2\dot{a}\dot{b} } {ab} +\frac{\dot{ b }^2}{b^2} &=& \kappa^2Y(E^2 + B^2) + 2 \frac{\ddot{b}}{b} + \frac{\ddot{a}}{a}  =\kappa^2 \rho\;, \label{denk1}
\\
\frac{2\ddot{b}}{b} + \frac{\dot{ b }^2 }{b^2}
&=&  \kappa^2Y(E^2 + B^2) + 2 \frac{\dot{a} \dot{b}}{a b} + \frac{\ddot{a}}{a} = -\kappa^2 p_x \;, \label{denk2}
\\
\frac{\ddot{a}}{a}  + \frac{\ddot{b}}{b} +  \frac{\dot{a}\dot{b} } {ab}  &=&
- \kappa^2Y(E^2 + B^2) +  \frac{\ddot{b}}{b} + \frac{\dot{a}\dot{b} } {ab} + \frac{\dot{ b }^2 }{b^2}  = - \kappa^2 p_y \label{denk3}
\end{eqnarray}
 where dot  means the derivative with respect to cosmic time $t$ and  we have used the condition (\ref{cond1})  as
\begin{eqnarray}\label{dif3}
   Y_R(E^2 - B^2) = \frac{1}{2\kappa^2} \, .
\end{eqnarray}
By subtracting  (\ref{denk1}) from (\ref{denk2}),  we arrive 
\begin{eqnarray}\label{denk12}
 \dot{a} \dot{b}  - a \ddot{b}  = 0 \;.
\end{eqnarray}
Equation (\ref{denk12}) gives the relation between the directional
scale functions as
\begin{eqnarray}
a = a_0\dot{ b} \label{at}
\end{eqnarray}
where $a_0$ arbitrary constant. 
By summing equations (\ref{denk1}) and (\ref{denk2}),
and using (\ref{denk12}) we obtain (\ref{denk3})
	or its the following equivalent form	
	\begin{eqnarray}
\frac{\ddot{a}}{a}  - \frac{ \dot{b}^2 }{ b^2}
+ \frac{\kappa^2 }{b^4}\left(  \frac{E_0^2}{Y} +YB_0^2\right) = 0
\  \, .\label{ab2}
\end{eqnarray}
We noticed that the derivative of equation (\ref{ab2}) gives the condition (\ref{dif3}). In order to solve the differential equation  (\ref{ab2}) we can choose the non-minimal function $Y(R)$ which gives the expansion function $b(t)$,  or alternatively we can choose the expansion function $b(t)$ which determines the non-minimal function $Y(R)$ in the Lagrangian, then $a(t)$ is determined by (\ref{at}). 
 In this study,   we  consider the second approach by taking  the  hyperbolic expansion. 
\begin{eqnarray}
b(t) = \sinh^k (\alpha t) \label{b}
\end{eqnarray}
where $\alpha$ and $k$ are  positive real numbers. Most of the recent observations indicate the presence of  the phase transition from deceleration to acceleration \cite{Perlmutter1999,Riess1998}.  The hyperbolic scale function (\ref{b}) leads to a deceleration parameter which can change sign from positive to negative. Furthermore,  it behaves  $b(t)\propto t^k$ in the beginning of the universe and $b(t) \propto e^{\alpha kt}$ in late times. Here the constant  $\alpha $ is  in unit $Gyr^{-1}$ and $\alpha k$ can be interpreted as  the Hubble parameter in late times. 
	
The scale function in the $x$ direction is obtained via (\ref{at}) as
\begin{eqnarray}
a(t) = a_0 \alpha k \sinh^{k-1}(\alpha t) \cosh(\alpha t) \;. \label{a21} 
\end{eqnarray}
When we look at  the limit     $	\lim\limits_{t \to \infty}   \frac{a(t)}{b(t)} = a_0\alpha k  $,  we see that we should set $a_0=\frac{1}{\alpha k}$,  
 in order to obtain isotropic case in late times. Thus $a(t)$ becomes 
  \begin{eqnarray}
  a(t) =  \sinh^{k-1}(\alpha t) \cosh(\alpha t) \;. \label{a2} 
  \end{eqnarray}
 By using equation (\ref{ab2}) we obtain the non-minimal function and the magnetic field as a solution of the model for $E=0$
\begin{eqnarray}
Y(t) &=&  \frac{(3k-2)\alpha^2 \sinh^{4k-2}(\alpha t)}{\kappa^2B_0^2} \;,  \label{Yt1} \\
B(t) &=& \frac{B_0}{\sinh^{2k}(\alpha t)} \label{B1} \;.
\end{eqnarray}
  The expansion functions (\ref{b}) and(\ref{a2}) also give  the following  solution with  the non-zero electric field and $B=0$ which is another model
\begin{eqnarray}
Y(t) = \frac{ \kappa^2 E_0^2  }{\alpha^2( 3k - 2) \sinh^{4k -2}(\alpha t)} \;, \label{Yt2} \\
E(t) =  \frac{(3k-2)\alpha^2\sinh^{2k-2}(\alpha t) }{E_0\kappa^2} \label{E1} \;.
\end{eqnarray} 

It is worthy to notice that the field equations (\ref{gfe1}) and (\ref{maxwell1}) or the differential equations (\ref{EB})-(\ref{ab2}) have the duality transformation given by $ (F, *F) \rightarrow (*YF, -YF)$ or $B\rightarrow -YE$, $B_0\rightarrow -E_0$ and $Y\rightarrow \frac{1}{Y}$ \cite{Sert2013MPLA}.
Thus each  model with magnetic field has a corresponding model with electric field and it can be found  by taking the duality transformation.
\begin{figure}[h]{}
	\centering
	\subfloat[ $k=1$   ]{   \includegraphics[width=0.4\textwidth]{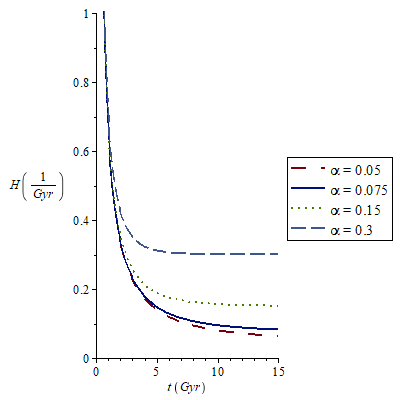}   }
	\subfloat[ $k= 2$ ]{ \includegraphics[width=0.4\textwidth]{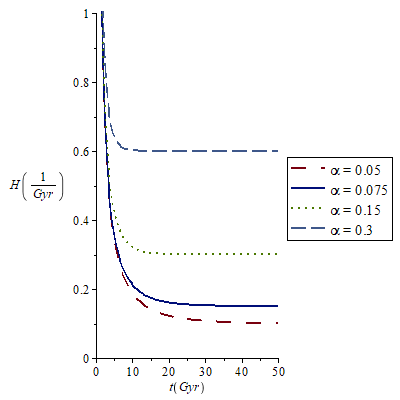}  } \\
	\parbox{6in}{\caption{{{\small{ The mean Hubble parameter $H(t)$ versus cosmic time $t$ 
						for various $\alpha$ values and 
						$k =1$, $k=2$,  respectively. }}}}}
\end{figure}

\begin{figure}[h]{}
	\centering
	\subfloat[ $\alpha= 0.08$   ]{   \includegraphics[width=0.4\textwidth]{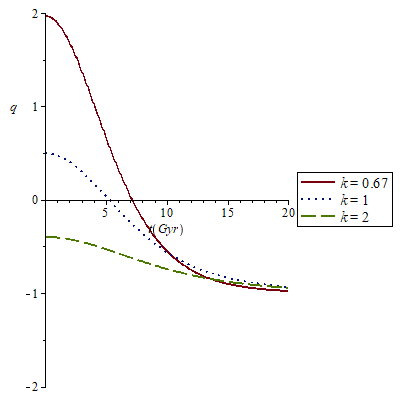}   }
	\subfloat[ $\alpha= 0.2$ ]{ \includegraphics[width=0.4\textwidth]{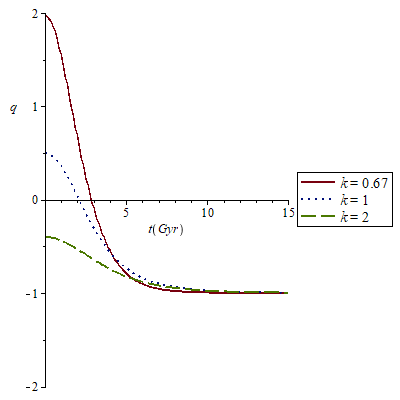}  }  \\
	\parbox{6in}{\caption{{{\small{ The deceleration parameter $q$ versus cosmic time $t$	for two $\alpha$ values and the corresponding  various $k$ values.}}}}}
\end{figure}

\begin{figure}[h]{}
	\centering
	\subfloat[ $k=1$   ]{   \includegraphics[width=0.4\textwidth]{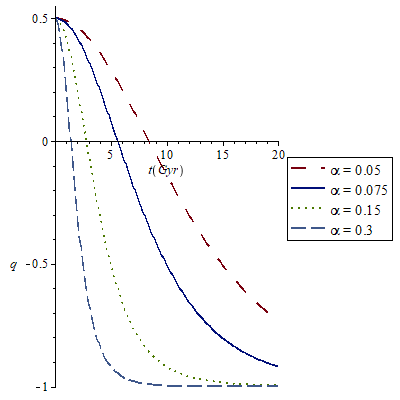}   }
	\subfloat[ $k= 2$ ]{ \includegraphics[width=0.4\textwidth]{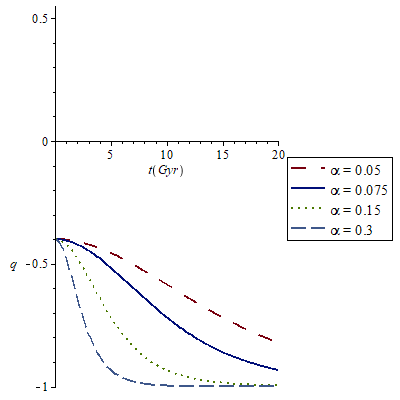}  }  \\
	\parbox{6in}{\caption{{{\small{ The deceleration parameter $q(t)$
	for two  $k$ values and the corresponding various  $\alpha$ values. }}}}}
\end{figure}

\begin{figure}[h]{}
	\centering
	\subfloat[ $k=1$   ]{   \includegraphics[width=0.4\textwidth]{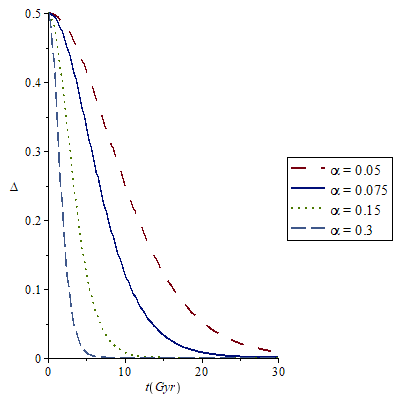}   }
	\subfloat[ $k= 2$ ]{ \includegraphics[width=0.4\textwidth]{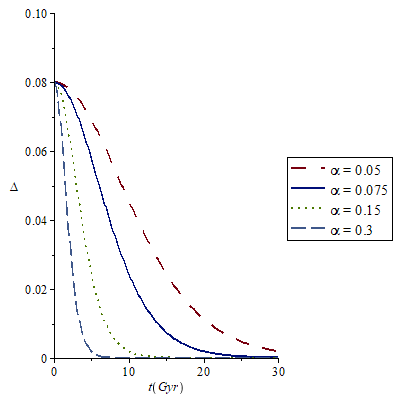}  } \\
	\parbox{6in}{\caption{{{\small{ The anisotropy parameter  $\Delta(t)$ 
	for various $\alpha$ values and $k =1$, $k=2$, respectively. }}}}}
\end{figure}

Then we calculate  the Ricci scalar for the expansion functions (\ref{b}) and (\ref{a2})
\begin{eqnarray}
R(t) = \frac{2\alpha^2(6k^2  -7k +2)}{\sinh^2(\alpha t)} + 12\alpha^2k^2 \;.
\end{eqnarray}
We see that the Ricci scalar is infinity at $t=0$ and it approaches the constant value $R = 12 \alpha^2 k^2 $, as $t\rightarrow \infty$. When we take the inverse function of  the Ricci scalar  and substitute it in  (\ref{Yt1})  and (\ref{Yt2}),  we obtain the non-minimal function $Y$  in terms of $R$.
Then we can write our model 
 \begin{eqnarray}\label{lag2}
L =  \frac{1}{2\kappa^2} R*1 - \frac{(3k-2)\alpha^2}{\kappa^2B_0^2}  \left(\frac{2\alpha^2(6k^2 -7k +2 )}{R-12\alpha^2k^2}\right)^{2k -1} F\wedge *F + \lambda_a  \wedge  T^a  ,
\end{eqnarray}
which admits the expansion functions (\ref{b}), (\ref{a2}) and the magnetic field (\ref{B1}) as a solution to the field equations.  Also, after the duality transformation \cite{Sert2013MPLA} we obtain  the corresponding model 
 \begin{eqnarray}\label{lag3}
L =  \frac{1}{2\kappa^2} R*1 -  \frac{\kappa^2E_0^2} {(3k-2)\alpha^2} \left(\frac{R-12\alpha^2k^2}{2\alpha^2(6k^2 -7k +2 )}\right)^{2k -1} F\wedge *F + \lambda_a  \wedge  T^a  ,
\end{eqnarray}
admitting the same expansion functions  (\ref{b}), (\ref{a2}) and the electric field (\ref{E1}) as a solution.
We note that the both models lead to the same scale factors and the cosmological parameters. Then we calculate  the following  parameters; the mean scale factor
\begin{eqnarray}
v= (ab^2)^{\frac{1}{3}}= (\cosh(\alpha t) \sinh^{3k-1}(\alpha t) )^{\frac{1}{3}} \;, 
\end{eqnarray}
the mean Hubble  parameter
\begin{eqnarray}\label{H1}
H= \frac{\dot{v}}{v} = \frac{ 2\alpha  (3k \cosh^2(\alpha t) - 1)}{3 \sinh(2\alpha t)} \;,
\end{eqnarray}
and the mean deceleration parameter
\begin{eqnarray}\label{q1}
q = -1 + \frac{d}{dt}(\frac{1}{H})   =  \frac{3(3k \cosh^2(\alpha t) - 2\cosh^2(\alpha t) +1) }{(3k \cosh^2(\alpha t) -1)^2} -1 \;.
\end{eqnarray}
 We demonstrate the behaviors of the mean Hubble parameter in Figure 1 and the mean   deceleration parameter $q$ in Figure 2 and Figure 3 for various parameter values.
  As we see from Figure 1, the mean Hubble parameter (\ref{H1}) goes to infinity as $t \rightarrow 0$. Furthermore, it  is a decreasing function of cosmic time and it approaches the constant value $H = k\alpha $ in late times.   Additionally,  Figure 2 shows that as  $\alpha $  values increase,  $q$ approaches $-1$ faster. 
We see  from (\ref{q1}) and the figures that the deceleration parameter $q$ is a monotonically decreasing function of cosmic time.  It starts from    $q(0) = \frac{4-3k}{3k -1 } $, at $t=0$, and decreases to $q=-1$ as $t\rightarrow \infty$. Then 
the phase transition from  deceleration to acceleration  occurs only when the initial deceleration parameter positive,  $q(0) = \frac{4-3k}{3k -1 } > 0$. Therefore $k $ must be   in the  interval $\frac{1}{3}<k < \frac{4}{3}$ for the phase transition. In other cases the universe expands  always in the  accelerated phase without phase transition.

The directional components of the parameters  are obtained as
\begin{eqnarray}
H_x =\frac{\dot{a}}{a} =\frac{ 2\alpha (k \cosh^2(\alpha t) -1 )}{\sinh(2\alpha t)} \;,\hskip 1.5 cm H_{y,z} = \frac{\dot{b}}{b} = k\alpha \coth(\alpha t) \;,
\end{eqnarray}
\begin{eqnarray}
q_x = -1 +\frac{d}{dt}\left(\frac{1}{H_x}\right) = \frac{(k-2) \cosh^2(\alpha t) +1}{(k \cosh^2(\alpha t) -1)^2} -1 \;, \hskip 0.5cm q_{y,z} = -1   +\frac{d}{dt}\left(\frac{1}{H_y}\right)= \frac{1- k \cosh^2(\alpha t)}{k \cosh^2(\alpha t)} \;.
\end{eqnarray}
The anisotropy parameter $\Delta$  and the shear scalar $\sigma^2$ are obtained by 

\begin{eqnarray}
\Delta &=& \frac{1}{3} \sum_{i=x,y,z} \left(\frac{H_i-H}{H}\right)^2 = \frac{2}{(3k \cosh^2(\alpha t) -1)^2} ,\;  \label{delta1}
\\ \sigma^2 &=&  \frac{1}{2}\sum_{i=x,y,z} (H_i -H)^2 = \frac{4\alpha^2}{3\sinh^2(2\alpha t)} \; .
\end{eqnarray}
Then we calculate the effective energy density and pressures 
\begin{eqnarray}
\rho =  \frac{k \alpha^2(3k \cosh^2(\alpha t) - 2 )}{\kappa^2\sinh^2(\alpha t)}\;, \hskip 1 cm p_x= -\rho \;, \hskip 1 cm p_y=p_z= \frac{\alpha^2( 5k -2 -3k^2\cosh^2(\alpha t) )}{\kappa^2 \sinh^2(\alpha t)} \; \label{rhop}
\end{eqnarray}
which lead to the equation of state 
\begin{eqnarray}
w_x= \frac{p_x}{\rho} = -1 \;, \hskip 1.5 cm w_y =w_z = \frac{p_y}{\rho}=  - \frac{3k^2\cosh^2(\alpha t) -5k +2 }{k( 3k \cosh^2(\alpha t) -2 )}\;.
\end{eqnarray}
We note that the positive effective  energy density condition of the model requires that $k>\frac{2}{3}$ from (\ref{rhop}).
We  see that the model has the  Big Bang  singularity at the beginning of the universe for $k>1$, since 
$	\lim\limits_{t \to 0}   H_{x,y,z} = \infty  $,  \ 
$	\lim\limits_{t \to 0}  \rho = \infty  $ and $	\lim\limits_{t \to 0}  a,b = 0  $.
Furthermore, we also see    $	\lim\limits_{t \to \infty}   \Delta = 0  $ from (\ref{delta1})  and  $	\lim\limits_{t \to \infty}   \frac{a(t)}{b(t)} = 1  $ from (\ref{b}) and (\ref{a2}) which means that  the universe approaches   isotropy and homogeneity  at late-times for all  positive $k$ values, see Figure 4.

\subsection{The Model with $k=1$}
In order to  demonstrate features of the models,   we focus on the simple case with $k=1$ in which  the expansion function (\ref{b}) takes the form 
\begin{eqnarray}
b(t) = \sinh(\alpha t) \label{b2}\;
\end{eqnarray} then, the  equation (\ref{a2})  leads to
\begin{eqnarray}
a(t) =  \cosh(\alpha t) \;.  \label{a22} 
\end{eqnarray}
By using equation (\ref{ab2}) we obtain the non-minimal function and the magnetic field
as a solution with $E=0$ \begin{eqnarray}
Y(t) = \frac{ \alpha^2 \sinh^2(\alpha t) }{\kappa^2 B_0^2} \;,  \label{Yt12} \\
B(t) = \frac{B0}{\sinh^2(\alpha t)} \label{B12} \;.
\end{eqnarray}
After the duality transformation given by $ (F, *F) \rightarrow (*YF, -YF)$ or $B\rightarrow -YE$, $B_0\rightarrow -E_0$ and $Y\rightarrow \frac{1}{Y}$ \cite{Sert2013MPLA},
 we can obtain the corresponding solution with non-zero electric field and  $B=0$ as 
\begin{eqnarray}
Y(t) = \frac{ \kappa^2 E_0^2  }{ \alpha^2 \sinh^2(\alpha t)} \;,  \label{Yt22} \\
E(t) = - \frac{\alpha^2}{E_0\kappa^2} \label{E12} \;.
\end{eqnarray} 
for the same expansion  functions (\ref{b2}), (\ref{a22}). By using  the expansion functions,   we calculate  the Ricci scalar as
\begin{eqnarray}
R(t) = \frac{2\alpha^2}{\sinh^2(\alpha t)} + 12\alpha^2 \;.
\end{eqnarray}

When we take the inverse function of  the Ricci scalar  and substitute it in  (\ref{Yt12})  and (\ref{Yt22}),  we obtain the non-minimal function $Y$  in terms of $R$.
Then  our model for the non-zero magnetic field becomes
\begin{eqnarray}\label{lag22}
L =  \frac{1}{2\kappa^2} R*1 - \frac{2\alpha^4}{\kappa^2B_0^2(R-12\alpha^2)}F\wedge *F + \lambda_a  \wedge  T^a ,
\end{eqnarray}
which gives the solution (\ref{b2}), (\ref{a22}) and (\ref{B12}). After the duality transformation  the corresponding model for the non-zero electric field
\begin{eqnarray}\label{lag32}
L =  \frac{1}{2\kappa^2} R*1 - \frac{\kappa^2E_0^2(R-12\alpha^2)}{2\alpha^4} F\wedge *F + \lambda_a  \wedge  T^a  ,
\end{eqnarray}
gives the solution  (\ref{b2}), (\ref{a22}) and (\ref{E12}).
Both models lead to the same scale factors and the cosmological parameters. Then we calculate  the following mean scale factor
\begin{eqnarray}\label{v2}
v= (ab^2)^{\frac{1}{3}}= ( \cosh(\alpha t) \sinh^2(\alpha t) )^{\frac{1}{3}} \;, 
\end{eqnarray}
the mean Hubble  parameter
\begin{eqnarray}
H= \frac{\dot{v}}{v} = \frac{2\alpha (3\cosh^2(\alpha t) -1)}{3 \sinh(2\alpha t)} \;,
\end{eqnarray}
and the mean deceleration parameter
\begin{eqnarray}\label{q2}
q = -1 + \frac{d}{dt}(\frac{1}{H})    =  \frac{3(\cosh^2(\alpha t) +1)}{(3\cosh^2(\alpha t) -1)^2} -1 \;.
\end{eqnarray}
We see  form (\ref{q2}) and Figure 3a that  the phase transition from deceleration to acceleration  occurs in the case with $k=1$, because of $q(0) = \frac{1}{2 } > 0$  and $ \lim\limits_{t \to \infty}   q(t) \longrightarrow  -1  $.  It is clearly seen from the figures that the mean deceleration parameter changes sign after a certain time from the beginning which depends on the parameter $\alpha$ and it approaches the value $-1$, monotonically. 

In order to obtain the mean  deceleration parameter  in terms of the  redshift $z = -1 + \frac{v_0}{v}$, we isolate $\cosh(\alpha t) $ from (\ref{v2}) as 
\begin{eqnarray}\label{cosht}
\cosh(\alpha t) = \frac{X}{6} +\frac{2}{X} 
\end{eqnarray}
where $ X= (  \ 108v^3 +12\sqrt{81 v^6 -12}\ \  )^{1/3}$,\ \ $v= \frac{v_0}{1+z}$ and $v_0$ present value of the scale factor. After   substituting  (\ref{cosht}) in (\ref{q2}), we obtain
\begin{eqnarray}
q(z) = \frac{\frac{X^2}{12} +\frac{12}{X^2} +5}{(\frac{X^2}{12} +\frac{12}{X^2} +1)^2 } -1\;.
\end{eqnarray}
The directional components of the parameters  are obtained as
\begin{eqnarray}
H_x =\frac{\dot{a}}{a} = \alpha \tanh(\alpha t) \;,\hskip 1.5 cm H_{y,z} = \frac{\dot{b}}{b} = \alpha \coth(\alpha t) \;,
\end{eqnarray}
\begin{eqnarray}
q_x = -1 +\frac{d}{dt}(\frac{1}{H_x}) = - \coth^2(\alpha t) \;, \hskip 1.5 cm q_{y,z} = -1  + \frac{d}{dt}(\frac{1}{H_y})= -\tanh^2(\alpha t)  \;.
\end{eqnarray}
The anisotropy parameter $\Delta$  and the shear scalar $\sigma^2$ are obtained by 
\begin{eqnarray}\label{Delta2}
\Delta = \frac{2}{(3\cosh^2(\alpha t) -1)^2} ,\;  \hskip 1.5 cm
\sigma^2 =  \frac{4\alpha^2}{3\sinh^2(2\alpha t)}\;.
\end{eqnarray}
Then we calculate the effective energy density and pressures 
\begin{eqnarray}
\rho =  \frac{\alpha^2(3\cosh^2(\alpha t) - 2 )}{\kappa^2\sinh^2(\alpha t)} \;, \hskip 1.5 cm p_x= -\rho \;, \hskip 1.5 cm p_y=p_z= -3\alpha^2
\end{eqnarray}
which leads to the equation of state 
\begin{eqnarray}
w_x= \frac{p_x}{\rho} = -1 \;, \hskip 1.5 cm w_y =w_z = \frac{p_y}{\rho}=  - \frac{3\sinh^2(\alpha t)}{3\cosh^2(\alpha t) -2}\;.
\end{eqnarray}

\begin{figure}[h]{}
	\centering
	\subfloat[ $w_x$ and $w_{y}=w_z$   ]{   \includegraphics[width=0.4\textwidth]{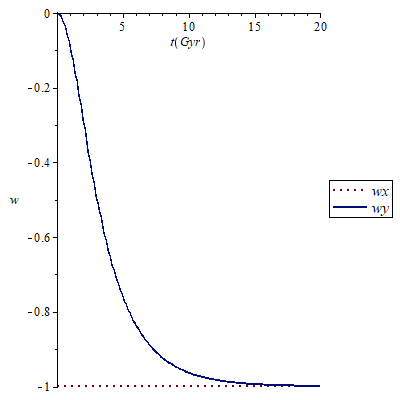}   }
	\subfloat[ $q$ ]{ \includegraphics[width=0.4\textwidth]{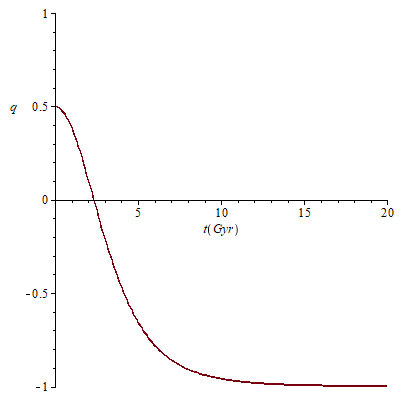}  }  \\
	\parbox{6in}{\caption{{{\small{ The directional equation of state parameters $w_x, \ w_{y,z}$ and deceleration parameter $q$ versus   cosmic time $t$ for $\alpha=0.08 $ .}}}}}
\end{figure}

\noindent We see from  equation (\ref{Delta2}) and Figure 4 that  the anisotropy parameter $\Delta $ goes to zero,  as t approaches to infinity,  	 $ \lim\limits_{t \to \infty}   \Delta = 0  $ and  $	\lim\limits_{t \to \infty}   \frac{a(t)}{b(t)} = 1  $ from (\ref{b}) and (\ref{a2}). It means that the universe becomes isotropic in late-times.   We also see that the model has a singularity at $t=0$.


\newpage
\section{Conclusions}

 \bigskip
\noindent We have investigated anisotropic cosmological solutions  of
the non-minimally coupled gravity in $Y(R)F^2$ form. After casting our model by a Lagrangian 4-form we have obtained
the variational field equations. Then we found 
solutions with only electric or magnetic fields  under the assumption of a spatially flat anisotropic space-time. We note that the anisotropy parameter  approaches to zero at late-times  and   $	\lim\limits_{t \to \infty}   \frac{a(t)}{b(t)} = 1  $. Therefore the universe becomes   homogeneous and  isotropic   at late-times.
 We also found that  the model has a singularity at $t=0$, since $ H\rightarrow \infty$, $\rho \rightarrow \infty $,    $b\rightarrow0$ and $v \rightarrow 0 $,  as   $t\rightarrow 0$, for all $\alpha$ when $k>\frac{2}{3}$.

Furthermore, as  seen  from the figures  and equation (\ref{q1}), the deceleration parameter $q$ starts from    $q(0) = \frac{4-3k}{3k -1 } $ and decreases to $q=-1$ as $t\rightarrow \infty$,    monotonically.   Additionally,  as the parameter  $\alpha$ increases it approaches $-1$ more faster.  Then 
the phase transition from  deceleration to acceleration  occurs only for  $\frac{1}{3}<k < \frac{4}{3}$. In other cases the universe expands  continuously in the accelerated phase without phase transition, this can be seen easily  in Figure 3b for some parameter values. The late-time acceleration can be realized by the constant curvature $R= 12\alpha^2 k^2 $ and the constant Hubble parameter $H=\alpha k$,  as $t\rightarrow \infty$\;.

 The recent observations indicate that the current value of the deceleration parameter is negative and it can  take values in the interval  $0<q<-1$ \cite{Ade2016,Ade2014}. Additionally, the possible deviations from isotropy is predicted by the upper bound $\Delta \lesssim  10^{-4}$ \cite{Campanelli2011} for type Ia supernovae through  a model independent way.

By  taking $t=13.8 $ Gyr in the anisotropy parameter (\ref{Delta2}), we find $\alpha\gtrsim 0.18$ for $k=1$. 
Most of the recent models predict a phase transition from the early decelerated phase  to the late time accelerated phase.
In this case with $k=1$,   the phase transition  is realized 
at $t\lesssim 2.3$ Gyr, see Figure 5b.
Since  after  $ \sim 20 $ Gyr  the anisotropy parameter becomes very small, it can be accepted that   the universe almost  reaches the isotropic phases about at that time. The deceleration parameter also reaches the value $-1$,  which corresponds to  de Sitter phase at the same time.
For the case with $k> 1$,  by using the same upper limit for the anisotropy parameter and (\ref{delta1}), we can obtain smaller   lower bound for $\alpha$ from  $\alpha \simeq 0.18$. In all of these cases the deceleration parameter takes values in the interval $-1<q<0$ for late times.
 Thus, we have given some limits on the free parameters
 in order that our model would exhibit a
behavior consistent with the current understanding of the 
universe.

 \section*{Acknowledgement}
 
 This study was supported by the Scientific Research Coordination Unit (BAP) of Pamukkale University.

\end{document}